\documentclass[useAMS,usenatbib]{mn2e}
\usepackage{graphicx}

\newcommand{\lsim}{\raisebox{-0.3ex}{\mbox{$\stackrel{<}{_\sim} \,$}}}

\title[Constraining neutron star EoS models]{Ways to constrain neutron star equation of state models using relativistic disc lines}
\author[Bhattacharyya S.]{Sudip Bhattacharyya\thanks{E-mail:
sudip@tifr.res.in}\\
Department of Astronomy and Astrophysics, Tata Institute
of Fundamental Research, Mumbai 400005, India}

\begin{document}

\date{
}

\pagerange{\pageref{firstpage}--\pageref{lastpage}} \pubyear{2009}

\maketitle

\label{firstpage}
\begin{abstract}
Relativistic spectral lines from the accretion disc 
of a neutron star low-mass X-ray binary can be modelled to infer 
the disc inner edge radius. A small value of this radius tentatively implies
that the disc terminates either at the neutron star hard surface, or 
at the innermost stable circular orbit (ISCO). Therefore an inferred disc inner
edge radius either provides the stellar radius, or can directly constrain
stellar equation of state (EoS) models using the theoretically computed 
ISCO radius for the spacetime of
a rapidly spinning neutron star. However, this procedure requires numerical
computation of stellar and ISCO radii for various EoS models and neutron star
configurations using an appropriate rapidly spinning stellar spacetime.
We have fully general relativistically
calculated about 16000 stable neutron star structures to explore
and establish the above mentioned procedure, and to show that the Kerr
spacetime is inadequate for this purpose. Our work 
systematically studies the methods to constrain EoS models using
relativistic disc lines, and will motivate future X-ray astronomy instruments.
\end{abstract}


\section{Introduction}\label{Introduction}

The core density of a neutron star is typically $5-10$ times higher than the
nuclear density. Such a dense matter at a relatively low temperature (e.g., $\sim 10^8$~K) cannot
be probed by heavy-nuclei collision experiments or with observations of the
early universe (\citet{LattimerPrakash2007} and references therein). Plausibly
the only way to probe this degenerate matter is to constrain the theoretically
proposed equation of state (EoS) models of neutron star cores \citep{ShapiroTeukolsky1983}.
For a given EoS model and an assumed central density, the stable structure of a 
non-spinning neutron star can be computed by solving the Tolman-Oppenheimer-Volkoff
(TOV) equation \citep{ShapiroTeukolsky1983}. These stable stars trace a single
curve in the mass ($M$) $-$ equatorial-radius ($R$) space. Therefore, measurement of mass
and radius of the same non-spinning neutron star would constrain the EoS models.
If the neutron star spins, then the EoS models can be constrained by the 
reliable measurements
of three independent parameters, such as mass, radius and spin-frequency ($\nu_{\rm spin}$)
of the {\it same} neutron star (e.g., Fig. 1 of \citet{Bhattacharyya2010}).
This is extremely difficult because of a number
of unknown systematics. Low-mass X-ray binaries (LMXBs)
can be particularly promising systems for such measurements, 
because several complementary methods are
available for them. A recent review by \citet{Bhattacharyya2010} describes
how the simultaneous application of these methods can reduce the systematic 
uncertainties.

One such method involves broad relativistic spectral lines from the inner portion of
the accretion disc. The strongest among these fluorescent spectral emission lines is the one for
the $n = 2 \rightarrow n = 1$ transition of the iron atom (or ion), and is 
observed from many
accreting supermassive and stellar-mass black hole systems (see \cite{ReynoldsNowak2003, 
Miller2007}; and references therein). Recently, 
\citet{BhattacharyyaStrohmayer2007b} has, for the first time, established that the
broad iron lines from neutron star LMXBs also originate from the inner accretion discs.
This discovery was soon confirmed by \citet{Cackettetal2008a} using data from a different
satellite. After these initial reports, the inner disc origin of broad iron line has been 
confirmed for several other neutron star LMXBs (e.g., \citet{Pandeletal2008, DAietal2009,
Cackettetal2009a, Papittoetal2009, Reisetal2009, diSalvoetal2009, 
Iariaetal2009, Cackettetal2009b}).

The broad iron line is affected by the strong gravity and spin of the compact star
(neutron star or black hole).
Before discussing how this relativistic line can be used to constrain the neutron
star parameters, let us see how it infers the black hole spin. 
The line shape, especially the extent of the red wing, carries the signature of 
the disc inner edge radius in the unit of stellar mass ($r_{\rm in}c^2/GM$).
This is because the red wing is primarily affected by longitudinal Doppler effect 
due to the oribital motion of the disc matter, which broadens the line,
and gravitational redshift, which shifts the line towards lower energies. 
Since, for a black hole the disc can extend up to the
innermost stable circular orbit (ISCO), the black hole
spin parameter $j = Jc/GM^2$ ($J$: total angular momentum;
$M$: mass), which determines the ISCO location, 
can be estimated by fitting the line shape with an appropriate
relativistic model for Kerr spacetime \citep{Laor1991,BeckwithDone2004,
Dovciaketal2004, Miller2007}.


A neutron star system is usually more complex than a black hole syetem because
of the following reasons. (1) While a spinning black hole is defined by only two
parameters (mass and spin), and the spacetime around it is the Kerr spacetime
having analytical expressions, 
the structures and spacetimes of rapidly spinning neutron stars, usually
harboured by LMXBs, may have to be numerically calculated
from at least two parameters apart from the EoS model \citep{Cooketal1994}.
(2) Unlike a black hole, a neutron star has a hard surface which can have observable
effects. Therefore, for a neutron star, the disc may be terminated 
either by ISCO or by the stellar surface \citep{Thampanetal1999}. 
An example of the competition between these two effects has been shown in the
Fig. 1 of \citet{MillerLambCook1998} and 
Fig. 1 of \citet{Bhattacharyyaetal2000}, which demonstrate that usually
circumferential radius $r_{\rm in}$ of disc inner edge
first decreases and then increases with the increase
of stellar spin for a given EoS 
model and mass. This complicates the measurement 
of $Jc/GM^2$ and other stellar parameters using the inferred 
circumferential $r_{\rm in}c^2/GM$, 
if it is not known whether the disc terminates at ISCO or at the stellar surface.
Therefore, theoretical 
computations of $r_{\rm in}$ as a function of stellar parameters for various EoS
models are essential. Such computations will also be useful to determine if a 
measured $r_{\rm in}c^2/GM$ directly gives the neutron star radius-to-mass ratio $Rc^2/GM$.
Note that the computations of $r_{\rm in}$ and stellar parameters 
will involve the numerical calculations
of rapidly spinning neutron star structures, and hence Kerr spacetime cannot
be used. Moreover, in order to constrain the EoS models, directly measurable 
neutron star parameters (e.g., $Rc^2/GM$, $M$, $\nu_{\rm spin}$; see \citet{Bhattacharyya2010})
should be expressed as functions of $r_{\rm in}c^2/GM$. Since $Jc/GM^2$
cannot usually be measured directly, a $Jc/GM^2$ vs. $r_{\rm in}c^2/GM$
plot may not be very useful to constrain the EoS models. In this paper,
such plots (Figs~\ref{abymrinm1.4}, \ref{abymrinm2.0} and \ref{abymrinmfrac0.95})
have been shown for comparisons with Kerr spacetime and to gain insight
(\S~\ref{Results}).

The procedure of constraining EoS models using directly measurable parameters
vs. $r_{\rm in}c^2/GM$ relations can be utilized only if 
$r_{\rm in}c^2/GM \lsim 6$, because a much larger value of $r_{\rm in}c^2/GM$
might imply the truncation of the disc by other effects (see \S~\ref{Discussion}).
Particularly useful would be $r_{\rm in}c^2/GM < 6$, because such values will
confirm the effect of neutron star spin on a corotating disc. Therefore, a crucial
question is whether observations show that $r_{\rm in}c^2/GM \lsim 6$.
\citet{Cackettetal2009b} report that $r_{\rm in}c^2/GM$ values fall into a
small range of $6-15$ for most of the neutron star LMXBs with established relativistic
disc lines. Moreover, although these authors could not measure a value of
$r_{\rm in}c^2/GM$ less than 6 (as they used non-spinning, i.e., Schwarzschild spacetime),
many of their fitted $r_{\rm in}c^2/GM$ values across the sources pegged at the
lower limit 6. This happenned for both phenomenological and reflection models,
and even when the Compton broadening was taken into account \citep{Cackettetal2009b}.
Such pegged best-fit values strongly suggest that $r_{\rm in}c^2/GM < 6$ for many cases, 
which is supported by the disc line fitting with spinning Kerr spacetime models 
\citep{BhattacharyyaStrohmayer2007b, Papittoetal2009, Reisetal2009, DAietal2010}.
This provides a good motivation to theoretically explore the above mentioned
procedure to constrain the EoS models.

Since, in this paper, we have computed neutron star structures and the
corresponding ISCO locations, let us briefly review some of the previous
studies on this topic. \citet{KluzniakWagoner1985} computed ISCO location
with slow stellar spin approximation. \citet{Cooketal1994} numerically 
calculated rapidly spinning neutron star structures in full general relativity
using realistic EoS models. \citet{MillerLambCook1998} used these structure
calculations to compute ISCO locations for a few cases. 
\citet{Stergioulasetal1999} studied if strange stars in LMXBs could be
excluded using the orbital frequency at ISCO. The relevance of higher-order
multipoles on the ISCO location was analytically probed by
\citet{ShibataSasaki1998}. \citet{BertiStergioulas2004} computed 
stellar models, multipole moments and ISCO locations, and 
compared the ISCO results with the \citet{Mankoetal2000a,Mankoetal2000b}
results for analytic solution for spacetime around rapidly spinning
neutron stars, Kerr ISCO results and 
\citet{ShibataSasaki1998} results. The results of \citet{BertiStergioulas2004} 
show that higher multipole moments cause differences between rapidly spinning
neutron star ISCO and Kerr ISCO. \citet{Pachonetal2006} and \citet{Sanabria-Gomezetal2010}
studied the ISCO around a spinning magnetic neutron star. \citet{Abramowiczetal2003}
calculated circular geodesics in the Hartle-Thorne metric for slowly spinning 
neutron stars. \citet{Bertietal2005} found that Hartle-Thorne approximation
gives ISCO radii accurate to within 1\%. \citet{Bejgeretal2010} suggested 
approximate analytic expressions for circular orbits around
rapidly spinning neutron stars.

Disc lines have so far been used to measure the black hole spin (e.g., 
\citet{BrennemanReynolds2006}). It has also been proposed that these lines
could be used to put an upper limit on the neutron star radius 
\citep{BhattacharyyaStrohmayer2007b,Cackettetal2008a} 
and to constrain other stellar parameters \citep{Bhattacharyya2010}.
However, to the best of our knowledge, plausible methods to measure these parameters,
and hence to constrain the EoS models
in a systematic way using appropriate spacetimes have not been studied so far.
The aim of this paper is to explore these methods. We have done so 
by calculating a huge number of neutron star structures, and 
the corresponding ISCO locations, if such a location is outside the neutron star.
Although, a large set of papers reported
the studies on such structures and locations (as indicated in the previous paragraph),
none of these aimed to probe neutron star parameters using iron lines.
In \S~\ref{Methods}, \S~\ref{Results} and \S~\ref{Discussion}, 
we describe our method, give the results and provide a discussion respectively.
Note that, since the accretion disc is believed to be thin, we have not considered 
nonequatorial orbits in our calculations. We have also not considered counterrotating orbits, 
because that would imply $r_{\rm ISCO}c^2/GM > 6$, which could be easily
confused with truncations caused by other effects (see \S~\ref{Discussion}).

\section{Method}\label{Methods}

In this section, we briefly describe the procedure to compute rapidly spinning 
neutron star structures, and their equilibrium sequences. The spacetime
around such a star can be described by the following metric (using $c=G=1$;
\citet{Bardeen1970, Cooketal1994}):
\begin{eqnarray}
{\rm d}s^2 = -{\rm e}^{\gamma+\rho}{\rm d}t^2 + {\rm e}^{2\alpha}({\rm d}r^2 + 
r^2{\rm d}\theta^2) + {\rm e}^{\gamma-\rho}r^2\sin^2\theta({\rm d}\phi 
- \omega{\rm d}t)^2,
\end{eqnarray}
where the metric potentials $\gamma$, $\rho$, $\alpha$, and the angular speed ($\omega$)
of the stellar fluid relative to the local inertial frame are all
functions of $r$ and $\theta$. 
For a given EoS model, and assumed values of stellar central density and 
polar-radius to equatorial-radius ratio, Einstein's field equations can be solved
to find out $r$ and $\theta$ dependence of $\gamma$, $\rho$, $\alpha$ and $\omega$,
as well as to obtain the stable stellar structure \citep{Cooketal1994, Dattaetal1998,
Bhattacharyyaetal2000, BhattacharyyaBhattacharyaThampan2001, BhattacharyyaMisraThampan2001,
BhattacharyyaThampanBombaci2001, Bhattacharyya2002}.
This equilibrium solution can then be used to calculate bulk structure parameters
(e.g., $M$, $R$, $J$) of the spinning neutron star.
Note that, henceforth, $R$ ($= r_e {\rm e}^{(\gamma_e-\rho_e)/2}$)
will denote equatorial circumferential radius of the neutron star
(equation B6 of \citet{Cooketal1994}).
The equations of motion of a test particle in the
spacetime around such a star are given in \citet{ThampanDatta1998}.
For example, the radial equation of motion is 
\.{r}$^2 \equiv {\rm e}^{2\alpha + \gamma + \rho}({\rm d}r/{\rm d}\tau)^2 = $\~{E}$^2 - 
$\~{V}$^2$, where, ${\rm d}\tau$ is the proper time, \~{E} is the specific 
energy and a constant of motion, and \~{V} is the effective potential. 
\~{V} is given by \~{V}$^2 = {\rm e}^{\gamma+\rho}[1 + 
\frac{l^2/r^2}{{\rm e}^{\gamma-\rho}}] + 2\omega$\~{E}$l - \omega^2 l^2$,
where $l$ is the specific angular momentum, a constant of motion.
We determine the radius of ISCO using the condition \~{V}$_{,rr}$ = 0,
where a comma followed by one $r$ represents a first-order
partial derivative with respect to $r$ and so on \citep{ThampanDatta1998}.
Thus we can compute both circumferential radius of ISCO (denoted by 
$r_{\rm ISCO}$) and $R$, and set the theoretical value of $r_{\rm in}$ to the
larger of them (see \S~\ref{Introduction}). 
For each EoS model, we have calculated $\nu_{\rm spin}$ sequences 
of equilibrium structures keeping $\nu_{\rm spin} =$ constant, and changing other
parameters. Since $\nu_{\rm spin}$ is not an input for computing structures,
a number of iterations are usually needed to compute the structure for a desired
$\nu_{\rm spin}$ value. The other sequences (e.g., $M$ sequence, $Rc^2/GM$ sequence;
figures of \S~\ref{Results})
have been calculated by interpolations of $\nu_{\rm spin}$ sequences.

We have used four representative EoS models of widely varying stiffness properties. This
ensures sufficient generality of our results. We briefly describe these models below.
{\it Model A} \citep{Sahuetal1993}: This very stiff EoS model with maximum non-spinning
mass $M_{\rm max} \approx 2.59 M_{\odot}$ is a field theoretical EoS for neutron-rich 
matter in beta equilibrium based on the chiral sigma model. {\it Model B} 
\citep{Akmaletal1998}: This stiff EoS model with $M_{\rm max} \approx 2.20 M_{\odot}$
is the Argonne $v_{18}$ model of two-nucleon interaction, with the three-nucleon 
interaction (Urbana IX [UIX] model) and the effect of relativistic boost corrections.
{\it Model C} \citep{Baldoetal1997}: This intermediate EoS model with $M_{\rm max} 
\approx 1.79 M_{\odot}$ is a microscopic EoS for asymmetric nuclear matter, 
derived from the Brueckner-Bethe-Goldstone many–body theory with explicit three-body terms.
{\it Model D} \citep{Pandharipande1971}: This very soft EoS model with $M_{\rm max}
\approx 1.41 M_{\odot}$ assumes an admixture of hyperons with the hyperonic potentials
similar to the nucleon-nucleon potentials, but altered suitably to represent the 
different isospin states.

\section{Results}\label{Results}

\begin{figure}
\centering
\hspace*{-0.85cm}
\includegraphics*[width=8.0cm,angle=0]{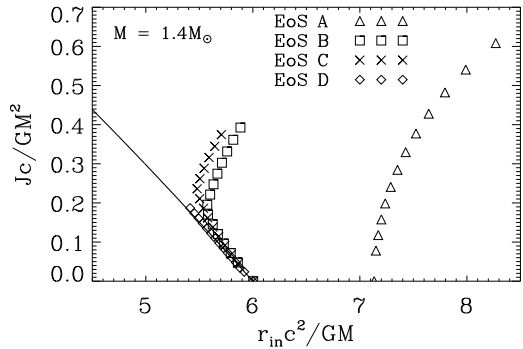}
\caption{Neutron star angular momentum parameter ($Jc/GM^2$) vs.
disc inner edge radius to stellar mass ratio ($r_{\rm in}c^2/GM$). The solid curve
is for Kerr spacetime (\S~\ref{Introduction}). Various symbols
give the curves for different neutron star EoS models (\S~\ref{Methods}) 
for $M = 1.4M_{\odot}$, and $\nu_{\rm spin}$ ranging from 0 Hz to 750 Hz. The
negative slope of the curves implies that the disc terminates at ISCO,
while the positive slope implies that it terminates at the stellar surface. This
figure shows that the realistic $r_{\rm in}c^2/GM$ values for neutron stars can
significantly deviate from the Kerr values.
\label{abymrinm1.4}}
\end{figure}

\begin{figure}
\centering
\hspace*{-0.85cm}
\includegraphics*[width=8.0cm,angle=0]{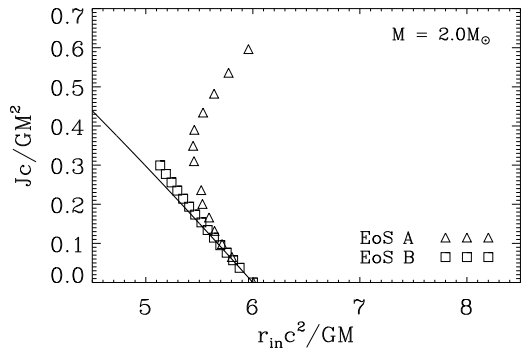}
\caption{Neutron star angular momentum parameter ($Jc/GM^2$) vs. 
disc inner edge radius to stellar mass ratio ($r_{\rm in}c^2/GM$).
Similar to Fig.~\ref{abymrinm1.4}, but for $M = 2.0M_{\odot}$. Note that 
stable neutron star structures do not exist for this high mass
for the softer equation of state models C and D (\S~\ref{Methods}).
\label{abymrinm2.0}}
\end{figure}

\begin{figure}
\centering
\hspace*{-0.85cm}
\includegraphics*[width=8.0cm,angle=0]{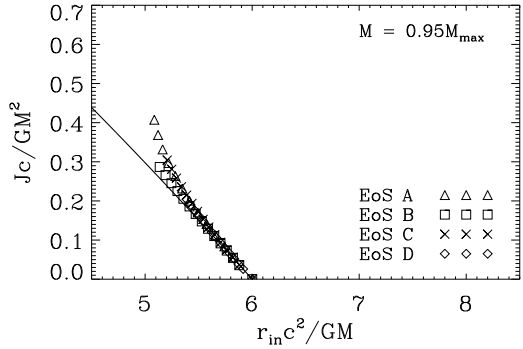}
\caption{Neutron star angular momentum parameter ($Jc/GM^2$) vs.
disc inner edge radius to stellar mass ratio ($r_{\rm in}c^2/GM$).
Similar to Fig.~\ref{abymrinm1.4}, but for $M = 0.95\times M_{\rm max}$. Here $M_{\rm max}$
is the maximum mass for a non-spinning stable neutron star for a 
given EoS. This figure shows that a fixed $M/M_{\rm max}$ gives
similar curves, and hence $M_{\rm max}$ may be a suitable parameter to
characterize an EoS model.
\label{abymrinmfrac0.95}}
\end{figure}

We have computed $\nu_{\rm spin}$ sequences (\S~\ref{Methods}) for 15 $\nu_{\rm spin}$
values in the range of $0-750$ Hz for each EoS model. About 16000 neutron star 
structures have been calculated to establish our results, and we give example figures 
in this section using a fraction of our computed numbers. The code to compute these
structures and $r_{\rm ISCO}$ values is well tested \citep{Dattaetal1998, ThampanDatta1998,
Bhattacharyyaetal2000, BhattacharyyaBhattacharyaThampan2001, BhattacharyyaMisraThampan2001,
BhattacharyyaThampanBombaci2001, Bhattacharyya2002}.
So far Kerr spacetime has been used to model the iron lines from spinning neutron star systems. 
Therefore, first we examine how our $Jc/GM^2$ vs. $r_{\rm in}c^2/GM$ plot deviates from 
the corresponding Kerr curve, in order to find out if Kerr calculations for 
iron lines can give acceptable constraints on neutron star parameters.
Fig.~\ref{abymrinm1.4} shows when the equatorial radius
of the neutron star is smaller than the ISCO radius, the deviation is relatively
small, but is not negligible, depending on the values of $\nu_{\rm spin}$ and $M$
and the chosen EoS model. But when the neutron star equatorial radius is
larger than the ISCO radius (see, for example, \citet{MillerLambCook1998}),
the deviation is large because stellar equatorial radius increases with the increase of 
$\nu_{\rm spin}$, and such a situation does not occur for black holes
(Kerr spacetime). For example, for the EoS model A, the stellar equatorial radius is
greater than the ISCO radius for all $\nu_{\rm spin}$ values for $M = 1.4 M_\odot$,
and hence the deviation is always very large. However, for $M = 2.0 M_\odot$, 
the stellar equatorial radius is less than the ISCO radius for smaller values
of $\nu_{\rm spin}$ (Fig.~\ref{abymrinm2.0}). In this case, the curves for both
EoS models A and B are closer to the Kerr curve compared to these curves for
$M = 1.4 M_\odot$. However, for $M = 2.0 M_\odot$, stable neutron star structures
do not exist for EoS models C and D. Therefore, from Fig.~\ref{abymrinm1.4}
and Fig.~\ref{abymrinm2.0} we find that the deviation is more for
(1) higher $\nu_{\rm spin}$, (2) lower $M$, and (3) stiffer EoS models.
The first two effects can also be seen in Fig. 1 of \citet{MillerLambCook1998}.
These effects are expected for $r_{\rm in}c^2/GM = Rc^2/GM$, because all these three
points result in the increase of $R$.
Let us now try to understand these points for $r_{\rm in}c^2/GM = r_{\rm ISCO}c^2/GM$.
The first point is understandable, because for $\nu_{\rm spin} = 0$ the spacetime
outside even a neutron star is Schwarzschild, which is the non-spinning special
case of Kerr. The second point could be understood from the fact that for lower $M$ 
the neutron star is usually less compact (that is the hard surface is farther 
from the centre), causing its spacetime to deviate more from that of a black hole. 
The third point may be explained from the lesser stellar compactness for 
a stiffer EoS model for given $M$ and $\nu_{\rm spin}$.
However, we find that if $M/M_{\rm max}$ is kept fixed instead of $M$,
the differences among the $Jc/GM^2$ vs. $r_{\rm in}c^2/GM$ curves for various
EoS models are small (Fig.~\ref{abymrinmfrac0.95}). This indicates that
$M_{\rm max}$ may be a suitable parameter to characterize an EoS model.

\begin{figure}
\centering
\hspace*{-0.85cm}
\includegraphics*[width=8.0cm,angle=0]{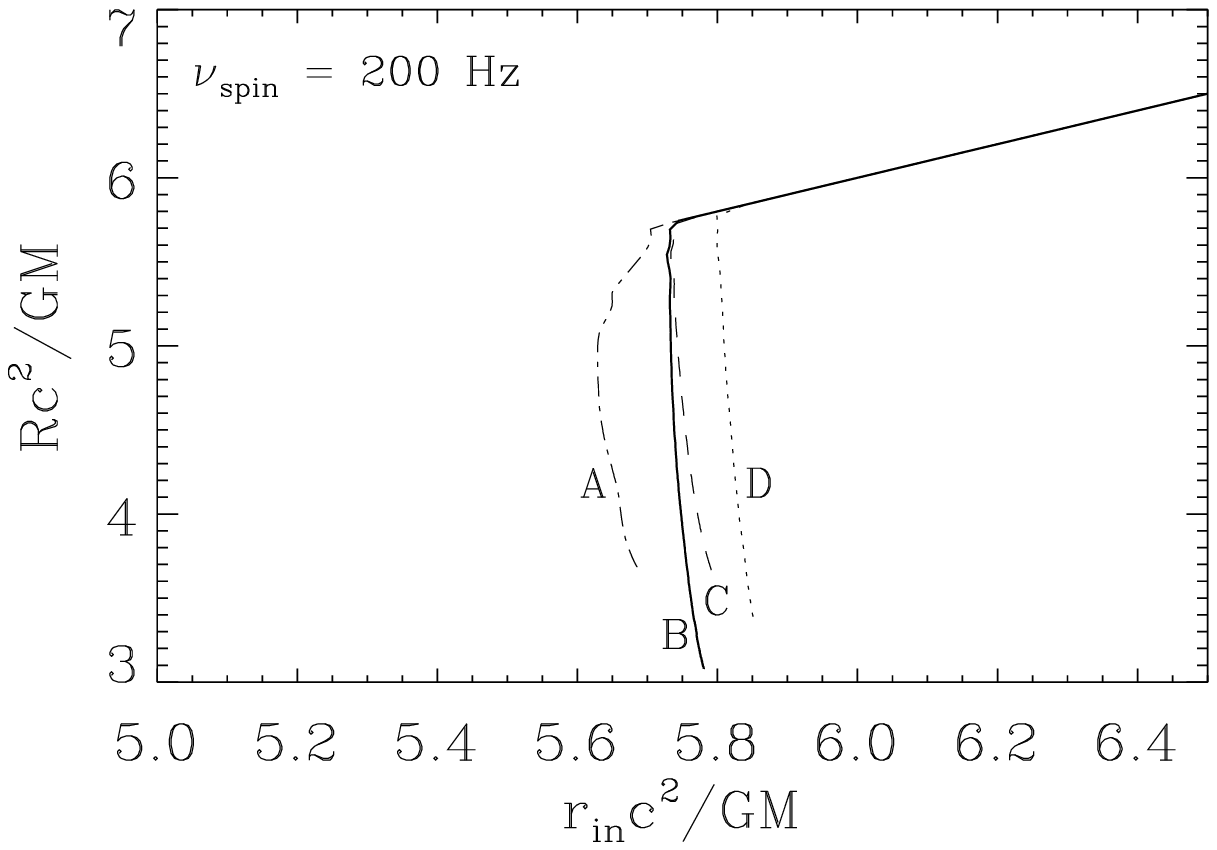}
\caption{Neutron star equatorial-radius-to-mass ratio ($Rc^2/GM$) vs. disc inner edge radius 
to stellar mass ratio ($r_{\rm in}c^2/GM$) for
various EoS models (\S~\ref{Methods}) for $\nu_{\rm spin} = 200$ Hz. 
Note that the oblique straight line portions of the curves are
for $r_{\rm in}c^2/GM = Rc^2/GM$. This
figure shows how a $r_{\rm in}c^2/GM$ value inferred from iron line can be
used to constrain the EoS models for a known $\nu_{\rm spin}$ value (\S~\ref{Results}).
\label{rbymrinnu200}}
\end{figure}

\begin{figure}
\centering
\hspace*{-0.85cm}
\includegraphics*[width=8.0cm,angle=0]{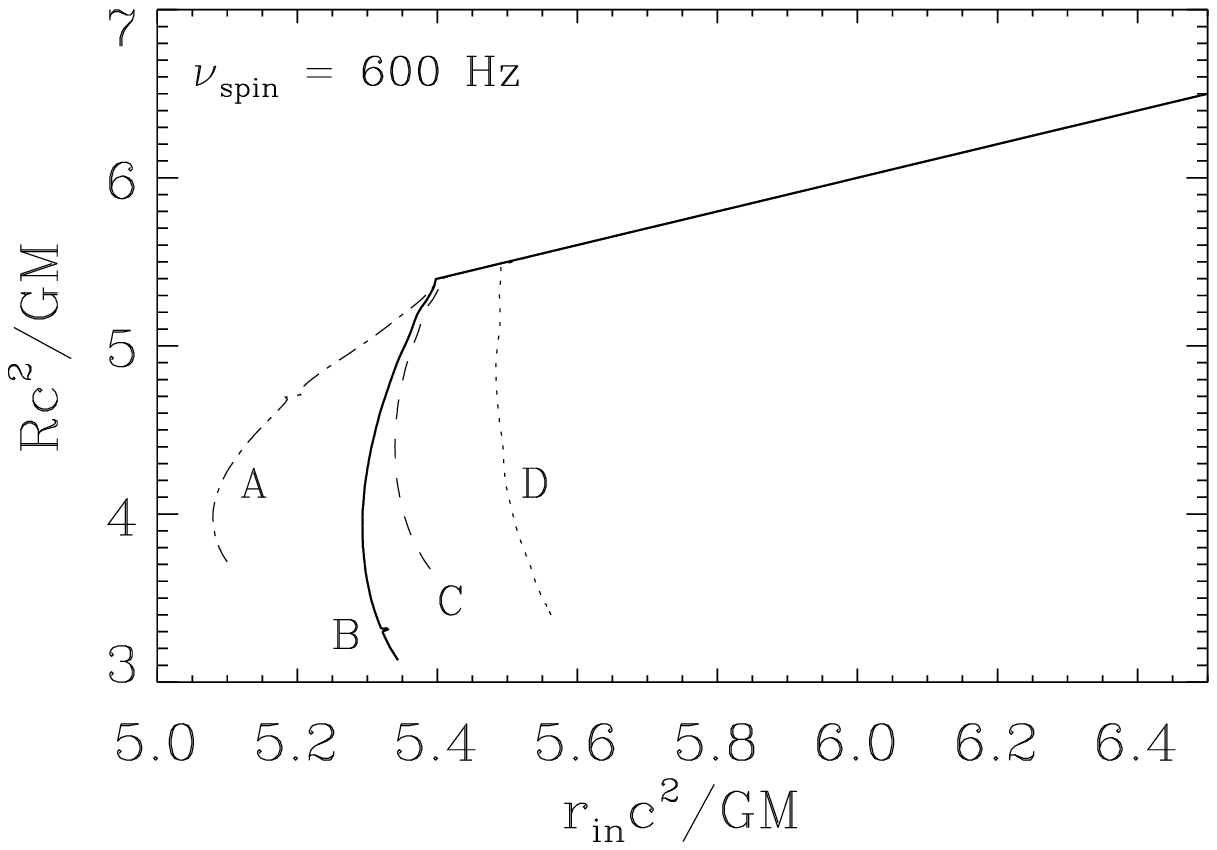}
\caption{Neutron star equatorial-radius-to-mass ratio ($Rc^2/GM$) vs. 
disc inner edge radius to stellar mass ratio ($r_{\rm in}c^2/GM$) for
various EoS models (\S~\ref{Methods}) for $\nu_{\rm spin} = 600$ Hz
(similar to Fig.~\ref{rbymrinnu200}).
\label{rbymrinnu600}}
\end{figure}

\begin{figure}
\centering
\hspace*{-0.85cm}
\includegraphics*[width=8.0cm,angle=0]{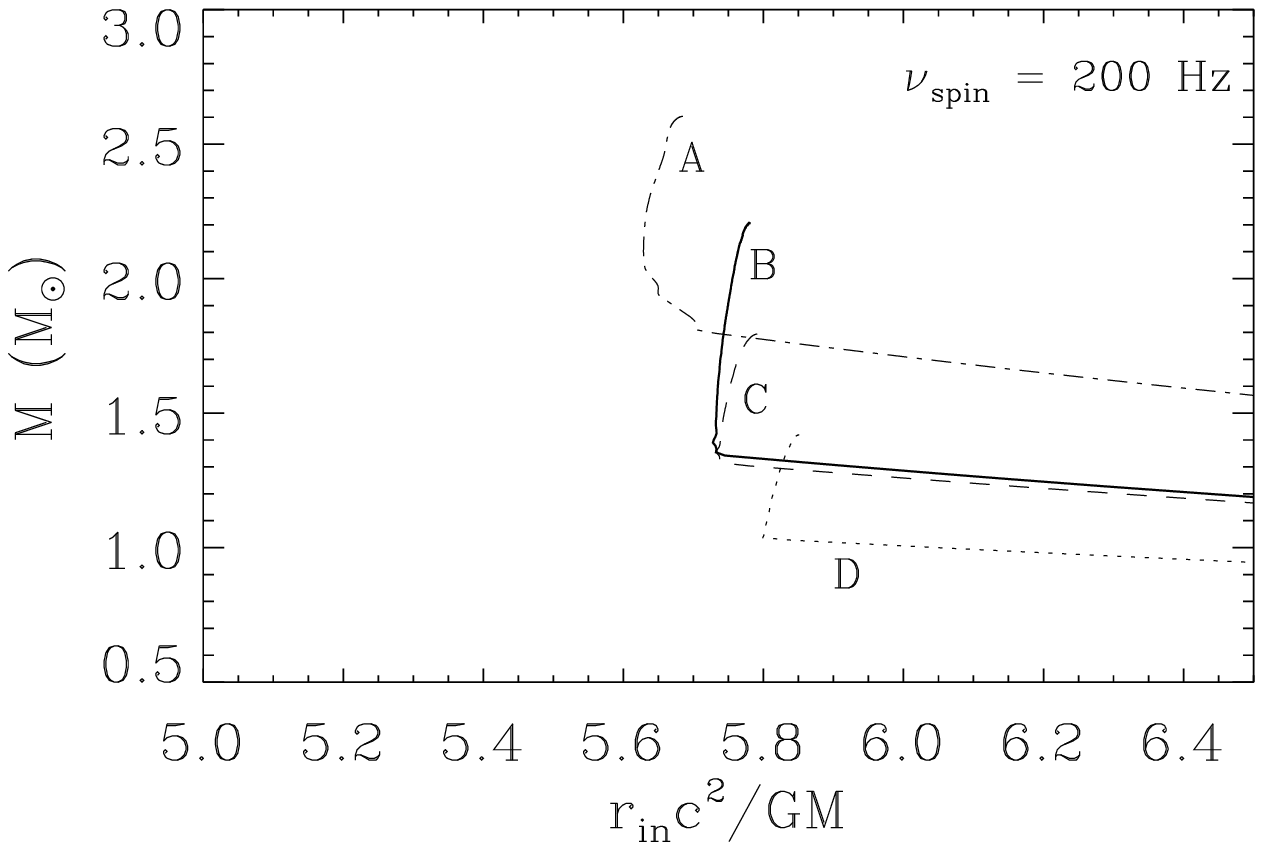}
\caption{Neutron star mass ($M$) vs. disc inner edge radius 
to stellar mass ratio ($r_{\rm in}c^2/GM$) for 
various EoS models (\S~\ref{Methods}) for $\nu_{\rm spin} = 200$ Hz. 
Note that the oblique straight line portions of the curves are
for $r_{\rm in}c^2/GM = Rc^2/GM$. This
figure shows how a $r_{\rm in}c^2/GM$ value inferred from iron line can be
used to constrain the EoS models for a known $\nu_{\rm spin}$ value (\S~\ref{Results}).
\label{mrinnu200}}
\end{figure}

\begin{figure}
\centering
\hspace*{-0.85cm}
\includegraphics*[width=8.0cm,angle=0]{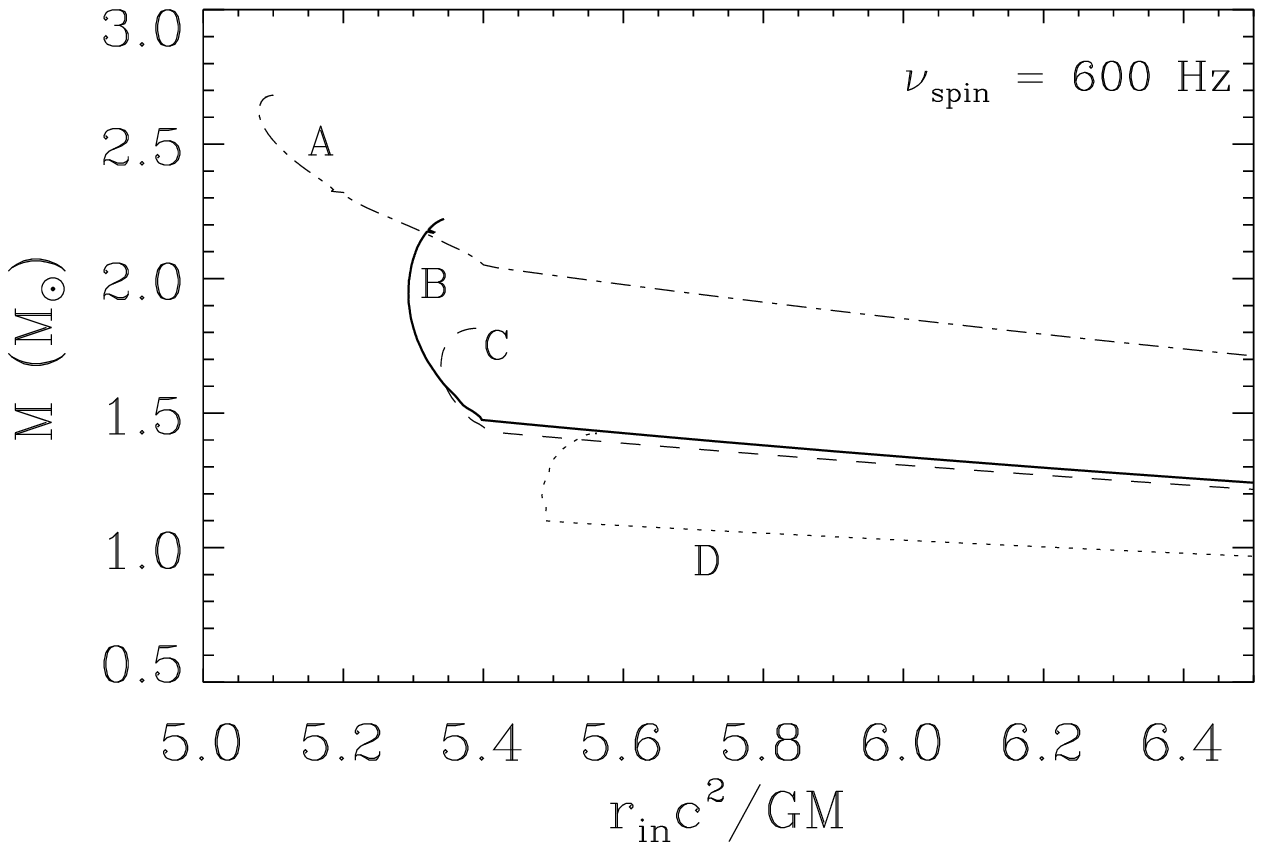}
\caption{Neutron star mass ($M$) vs. disc inner edge radius 
to stellar mass ratio ($r_{\rm in}c^2/GM$) for
various EoS models (\S~\ref{Methods}) for $\nu_{\rm spin} = 600$ Hz
(similar to Fig.~\ref{mrinnu200}).
\label{mrinnu600}}
\end{figure}

\begin{figure}
\centering
\hspace*{-0.85cm}
\includegraphics*[width=8.0cm,angle=0]{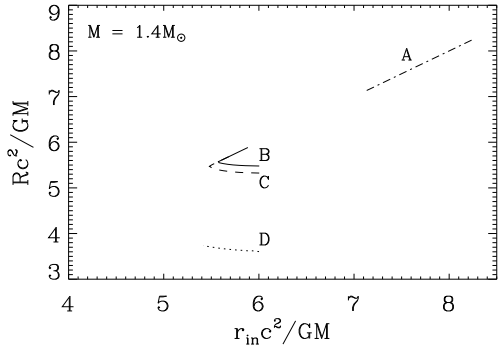}
\caption{Neutron star equatorial-radius-to-mass ratio ($Rc^2/GM$) vs. disc inner edge radius
to stellar mass ratio ($r_{\rm in}c^2/GM$) for various EoS models (\S~\ref{Methods}) for 
$M = 1.4M_{\odot}$. Note that the oblique straight line portions of the curves are
for $r_{\rm in}c^2/GM = Rc^2/GM$.
This figure shows, for a known $M$, how a $r_{\rm in}c^2/GM$ value inferred 
from iron line can be used to constrain the EoS models (\S~\ref{Results}).
\label{rbymrinm1.4}}
\end{figure}

\begin{figure}
\centering
\hspace*{-0.85cm}
\includegraphics*[width=8.0cm,angle=0]{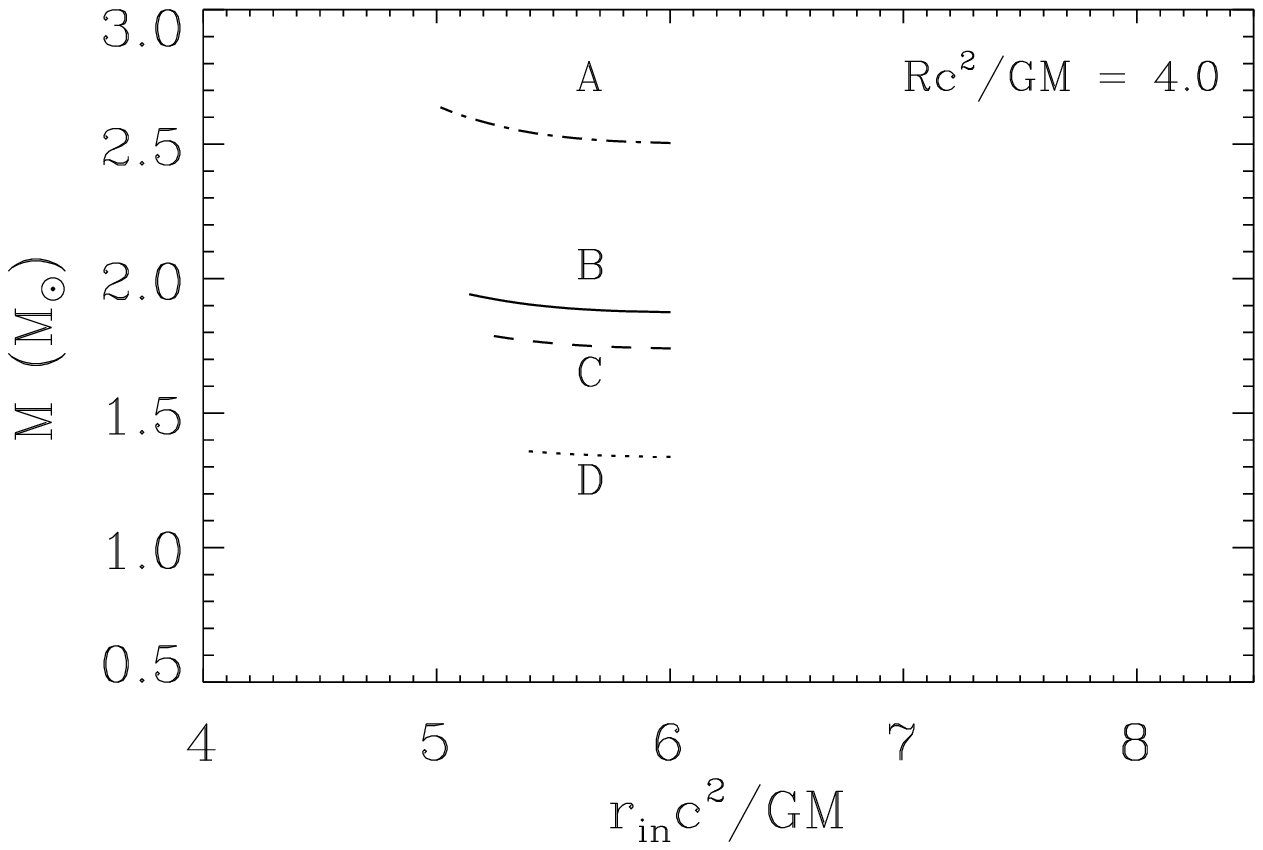}
\caption{Neutron star mass ($M$) vs. disc inner edge radius
to stellar mass ratio ($r_{\rm in}c^2/GM$) for various EoS models (\S~\ref{Methods}) for
$Rc^2/GM = 4.0$. This figure shows, for a known $Rc^2/GM$, 
how a $r_{\rm in}c^2/GM$ value inferred
from iron line can be used to constrain the EoS models (\S~\ref{Results}).
\label{mrinrbym4.0}}
\end{figure}

After finding that the Kerr spacetime is usually not good enough to model 
$r_{\rm in}c^2/GM$, and since this spacetime cannot be used for neutron star
parameter calculation, we have explored ways to constrain stellar EoS models 
using appropriate general relativistic computations for neutron stars 
(see \S~\ref{Methods}).
We have checked the relation between $r_{\rm in}c^2/GM$ and a few stellar parameters, 
which can be measured from independent methods. These parameters
are $\nu_{\rm spin}$, $Rc^2/GM$ and $M$. Whenever $\nu_{\rm spin}$ can be
measured (using regular pulsations or burst oscillations; \citet{Bhattacharyya2010}),
it is measured very accurately. Therefore, in the Figs.~\ref{rbymrinnu200},
\ref{rbymrinnu600}, \ref{mrinnu200} and \ref{mrinnu600}, we have fixed $\nu_{\rm spin}$.
$Rc^2/GM$ and $M$, on the other hand, may be constrained in a range 
(using thermonuclear X-ray bursts, binary orbital motions, etc.; 
\citet{Bhattacharyya2010}), and hence we have used them as dependent
variables in these figures. Figs.~\ref{rbymrinnu200} and \ref{rbymrinnu600}
show $Rc^2/GM$ vs. $r_{\rm in}c^2/GM$ plots for two values of $\nu_{\rm spin}$
and four EoS models. For a given $\nu_{\rm spin}$ value, and if the stellar
equatorial radius is less than the ISCO radius, each EoS model traces a distinct
curve. These curves, which are more separated from each other 
for higher $\nu_{\rm spin}$, can be
used to constrain EoS models from the $r_{\rm in}c^2/GM$ value inferred from
the iron line fitting, and the $Rc^2/GM$ value measured independently 
(see \citet{Bhattacharyya2010}). These figures show that even if $Rc^2/GM$
is not well constrained, a suitable upper limit of $r_{\rm in}c^2/GM$ can 
reject softer EoS models. If the stellar equatorial radius is larger than 
the ISCO radius, an oblique straight line is found for all EoS models,
and an inferred $r_{\rm in}c^2/GM$ value directly gives the $Rc^2/GM$ value.
Figs.~\ref{rbymrinnu200} and \ref{rbymrinnu600} clearly show the value
of $r_{\rm in}c^2/GM$ (for a given $\nu_{\rm spin}$),
above which $r_{\rm in}c^2/GM$ can be used to directly infer the 
EoS-model-independent $Rc^2/GM$ value.
Figs.~\ref{mrinnu200} and \ref{mrinnu600} show even if, instead of
$Rc^2/GM$, $M$ is known from independent measurement \citep{Bhattacharyya2010},
$r_{\rm in}c^2/GM$ inferred from iron line can be used to constrain the EoS models.
Even for an unknown $M$, a suitable upper limit of $r_{\rm in}c^2/GM$ can
be used to reject softer EoS models.
Fig.~\ref{rbymrinm1.4} explores how for a known $M$ value, $r_{\rm in}c^2/GM$
inferred from iron line and independently constrained $Rc^2/GM$ can be used
to constrain EoS models. Similar result is shown in Fig.~\ref{mrinrbym4.0},
where $Rc^2/GM$ is known and $M$ is reasonably constrained. However, 
for these two procedures (Figs.~\ref{rbymrinm1.4} and \ref{mrinrbym4.0}),
measurements of $Rc^2/GM$ and $M$ appear to be more important than an inferred
$r_{\rm in}c^2/GM$.

\section{Discussion}\label{Discussion}

In this paper, we explore ways to constrain neutron star EoS models 
by comparing the inferred values of $r_{\rm in}c^2/GM$ with the theoretical values. 
A $r_{\rm in}c^2/GM$ value may be inferred by fitting the relativistic disc lines with
appropriate models \citep{Bhattacharyya2010}.
We have shown that $r_{\rm in}c^2/GM$ calculated from Kerr spacetime 
is inadequate to distinguish between EoS models even when $r_{\rm in}c^2/GM =
r_{\rm ISCO}c^2/GM$, and can largely differ from the correct value
for rapidly spinning neutron star spacetime when $r_{\rm in}c^2/GM =
Rc^2/GM$. We have numerically computed a huge number
of $r_{\rm in}c^2/GM$ values, assuming that the disc terminates
either at ISCO or at the stellar hard surface. This should be at least 
approximately true for $r_{\rm in}c^2/GM \lsim 6$, although systematics may be introduced 
due to the effects of magnetic field and radiative pressure.
Such systematics would imply that any inferred $r_{\rm in}c^2/GM$ value is
basically an upper limit of $r_{\rm ISCO}c^2/GM$ or $Rc^2/GM$, 
which can still be used to reject softer EoS models
(Figs.~\ref{rbymrinnu200}, \ref{rbymrinnu600}, \ref{mrinnu200} and \ref{mrinnu600}).
These systematics can be reduced by detailed modelling of these effects, 
as well as by independent observations. We have studied the relations between 
$r_{\rm in}c^2/GM$ and several directly measurable neutron star parameters
\citep{Bhattacharyya2010}, in order to establish new ways to constrain EoS models. 
We have found that this iron line method could be effective to constrain 
EoS models, if the neutron star spin frequency is independently measured
\citep{Bhattacharyya2010}.
This work is timely and important, as it provides motivation for future
X-ray missions, and because of the rapid 
progress in the disc line field via observations with {\it XMM-Newton}, 
{\it Suzaku} and {\it Chandra}.

\section*{Acknowledgments}

We thank A. Thampan for the rapidly spinning neutron star structure computation code,
and A. Gopakumar and B. Iyer for discussion.
This work was supported in part by US NSF grant AST 0708424.

\end{document}